# Resilient Microgrid Formation Considering Communication Interruptions

Jian Zhong, *Student Member, IEEE*, Chen Chen, *Senior Member, IEEE*, Young-Jin Kim, *Senior Member, IEEE*, Yuxiong Huang, *Member, IEEE*, Mengjie Teng, *Student Member, IEEE*, Yiheng Bian, *Student Member, IEEE*, and Zhaohong Bie, *Fellow, IEEE*

***Abstract*—Distribution system (DS) communication failures following extreme events often degrade monitoring and control functions, thus preventing the acquisition of complete global DS component state information, on which existing post-disaster DS restoration methods are based. This letter proposes methods of inferring the states of DS components in the case of incomplete component state information. By using the known DS information, the operating states of unobservable DS branches and buses can be inferred, providing complete information for DS performance restoration before full communication recovery.**

## I. Introduction

DISTRIBUTION system (DS) resilience involves the use of various resources to quickly restore power to consumers following a disaster [1]. A key method of DS restoration is to control the automated switches to change the DS topology, thereby forming microgrids with distributed generators (DGs) supplying power to loads [2]. Notably, the control of automated switches in DS restoration is inseparable from the support of communication networks [3]. If communication between the feeder terminal units (FTUs) and the operation center (OC) is interrupted due to communication equipment or link failure, the monitoring and control functions of feeder automation (FA) to the relevant buses and branches will also be lost [4]. Therefore, as an important factor that increase the vulnerability of DSs, communication interruption must be considered in FA. However, most existing DS restoration studies assume that the global states of DS buses and branches are known, which is based on the premise that the DS cyber sectors are intact, allowing for the rapid development of restoration schemes after extreme events [2], [5]. Nevertheless, global observability of DSs in communication interruption events is an ideal situation that is difficult to achieve. Therefore, the unknown state limits the validity of the radiality constraints on which mainstream microgrid formation algorithms are based. Some other studies exclude the non-loss-of-power and loss-of-observability components from the DS restoration optimization, which may be overly conservative, making load recovery ineffective.

The impacts of communication failures on post-disaster DS restoration have been recognized in several existing studies. The use of mobile base stations [3], unmanned aerial vehicle base stations [4], and manual maintenance to restore communication between the OC and FTUs has been discussed. In any case, DS restoration cannot be performed before full communication recovery in the above studies. This recovery process takes considerable time, especially when there is equipment failure or extensive communication breakdown. Determining how to perform efficient DS restoration before full communication recovery is a practical problem in reducing the impacts of blackouts.

To fill the gap in this area, this letter proposes new communication interruption processing methods. Utilizing the available DS information, these methods are able to infer the unobservable DS component states when the monitoring function is degraded. With the help of such inference, control schemes can be developed for DS restoration. To our knowledge, this work is the first DS restoration method with an incomplete monitoring function.

## II. Fundamentals

This letter focuses on the processing of unobservable but electrified buses. To ensure safety, buses without observability and power supply cannot be connected to the DGs until communication with their FTUs is reestablished. When there is a communication failure between the FTU of bus $k$ and the OC as shown in Fig. 1, the monitoring and control functions for the corresponding loads, branches, and automated switches will be lost. The DS restoration scheme cannot be derived due to incomplete topology state information.

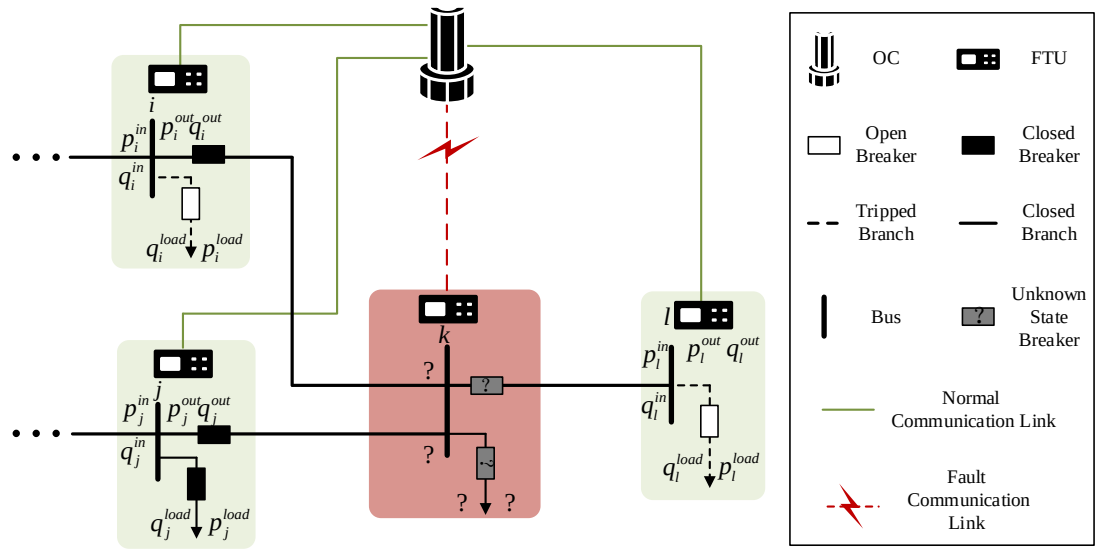

**Fig. 1.** Diagram of FTU communication interruption.

However, the FTUs of buses $j$ and $l$ communicate normally with the OC and provide effective information for unknown state evaluation. First, the power flowing into and out of buses $j$ and $l$ can then be used to infer the operating states of branches $k-l$ and bus $k$. Second, the paths from the electrified bus $l$ to power sources can likewise help infer the state of the branch $k-l$. Moreover, in cases of large-scale communication failures where some unobservable branch operating states are unable to be inferred by the above two methods, it is still possible to infer the states by performing certain disturbance actions, such as rapid load cut off and pick up, and then observing the power variation of DGs. The above methods are also types of topology state estimation within an unobservable local area. Nevertheless, unlike the traditional state estimation methods, they are a direct application to, and part of, DS load recovery schemes. After corresponding processing, loads can be recovered using the single-commodity flow (SCF) method [5].

## III. Methodology and Algorithm

For a DS with partial communication interrupted FTUs, its buses and branches can be divided into known-state sets $\mathcal{N}^k$ and $\mathcal{L}^k$ and unknown-state sets $\mathcal{N}^u$ and $\mathcal{L}^u$, respectively. The buses connected to DGs are sorted into a set $\mathcal{G}$. The unknown states of the components can be inferred in the following three ways.

*1) Power flow determination:* The state of an unknown branch $k=(i,\ j)$ can be inferred by the power flow data from the FTUs at both ends of the buses. Whether the active and reactive power is zero indicates whether the branch is open or not, which can be expressed as two inequality constraints:

$$(|p_{k,y}|+|q_{k,y}|)/M \le s_k^L \le (|p_{k,y}|+|q_{k,y}|)\cdot M, k=(i,\ j)\in \mathcal{L}^u, y\in\{i,j\}\&y\in\mathcal{N}^k, \quad (1)$$

$$0\le s_k^L \le s_i^N + s_j^N - 2, k\in\mathcal{L}^u, \quad (2)$$

where a very large number $M$ is used to express the conditional statement; $p_{k,i}$ and $q_{k,i}$ denote the active and reactive power, respectively, injected into bus $i$ through branch $k$; and $s_k^L$ is used to represent the state of branch $k$, with "1" indicating closed "0" indicating open, and "2" indicating unknown.

*2) Power supply path determination:* The unobservable states of branches can be evaluated by the paths from the DGs to the electrified buses. Adjacency matrices, $A^{d1}$ and $A^c$, can be generated by the Floyd algorithm under the assumption that the states of all unknown-state branches are disconnected or connected. The matrices are then used to count the paths from the buses to the DGs. If there is only one path from an electrified bus to the DGs, all branches on the path should be closed, which can be described as:

$$1-\left(r_i^{d1}+|1-r_i^c|+|1-s_i^N|\right)\cdot M \le s_k^L \le 1+\left(r_i^{d1}+|1-r_i^c|+|1-s_i^N|\right)\cdot M, i\in\mathcal{N}^k, \forall k\in\mathcal{P}_i^{d1}, \quad (3)$$

where $r_i^{d1}$ and $r_i^c$ are the numbers of paths from bus $i$ to DGs in matrices $A^{d1}$ and $A^c$, respectively; $\mathcal{P}_i^{d1}$ is the set of branches on the paths in matrixes $A^{d1}$; and $s_i^N$ is the state of bus $i$, with "0" indicating electrified, "1" indicating not electrified, and "2" indicating unknown.

*3) Power disturbance determination:* Some buses with critical loads have tie switches for the selection of multiple power supplies. In the case that the abovementioned area is not observable, if the loads of the observable and controllable buses at the lower end of the unobservable components are allowed to be cut off or picked up for a short period, the topological connections can be obtained by observing the value variation of the output power from the DGs, which can be formulated as:

$$\delta_i = r_i^c\cdot|1-r_i^c|, i\in\mathcal{N}^k, \quad (4)$$

$$\gamma_i = |1-s_i^N| + |[p_{b,x}^G - p_{a,x}^G - p_i^O]|, i\in\mathcal{N}^k, x\in\mathcal{G}, \quad (5)$$

$$\delta_i - \gamma_i\cdot r_i^{d1}\cdot M \le \delta_i\cdot s_k^L \le \delta_i + \gamma_i\cdot r_i^{d1}\cdot M, i\in\mathcal{N}^k, x\in\mathcal{G}, \forall k\in\mathcal{P}_{i,x}^{d1}, \quad (6)$$

where $[\alpha]$ represents taking an integer as the parameter $\alpha$; $\delta_i$ and $\gamma_i$ are auxiliary parameters for bus $i$; $p_i^O$ is the load that is cut off or picked up in the disturbance operation; $p_{b,x}^G$ and $p_{a,x}^G$ are the output power from DG $x$ before and after the operation, respectively; and $\mathcal{P}_{i,x}^d$ is a set of branches on the path from bus $i$ to DG $x$ in matrix $A^{d1}$.

*4) Remaining unknown state component handling:* For branches without a monitoring function, the control function is also disabled; i.e.,

$$b_k^L = 2, k\in\mathcal{L}^u, \quad (7)$$

where $b_k^L$ is a variable representing the opening or closing decision for branch $k$ in the subsequent DS restoration work, with "1" indicating closing, "0" indicating opening, and "2" indicating that no control is performed.

For buses that cannot derive power supply paths using the above methods, their subsequent branches are not allowed to be closed for load pickup; i.e., if the number of paths from an electrified bus to the DGs is zero when the remaining unknown branches are assumed disconnected, the branches connected to the bus are not allowed to operate in the restoration process, which can be written as follows:

$$2-\left(r_i^{d2}+|1-s_i^N|\right)\cdot M \le b_k^L \le 2+\left(r_i^{d2}+|1-s_i^N|\right)\cdot M, i\in\mathcal{N}^k, \forall k\in\mathcal{F}_i, \quad (8)$$

where $\mathcal{F}_i$ is the set of branches connected to bus $i$; $r_i^{d2}$ is the number of paths from bus $i$ to the DGs in the second adjacency matrix $A^{d2}$ generated by the Floyd algorithm under the assumption that all unknown state branches are disconnected.

The above processing methods can be summarized by Algorithm 1 when processing communication interruptions.

**Algorithm 1** unobservable state inference

**input** DS components states $\{s_i^N, s_k^L | i\in(\mathcal{N}^k,\mathcal{N}^u), k\in(\mathcal{L}^k,\mathcal{L}^u)\}$ and power flow data $\{p_{k,i} | i\in(\mathcal{N}^k,\mathcal{N}^u), k\in(\mathcal{L}^k,\mathcal{L}^u)\}$;
**function** $FloydProcess(\{s_k^L, k\in\mathcal{L}^k\},\{s_k^L, k\in\mathcal{L}^u\})$
  generate adjacency matrix $A$ by the Floyd Algorithm, then evaluate the path $\mathcal{P}_{i,x}$ from DG $x$ to bus $i$ and the number of paths $r_i$;
**output** $A$, $\mathcal{P}_i=\{\mathcal{P}_{i,x} | i\in(\mathcal{N}^k,\mathcal{N}^u), x\in\mathcal{G}\}$, $r_i$;
**for** $k=(i,j)\in\mathcal{L}^u$, $i\in\mathcal{N}^k$, **or** $j\in\mathcal{N}^k$, **do**
  **if** $|p_{k,i}|+|q_{k,i}|\ne 0$ **or** $|p_{k,j}|+|q_{k,j}|\ne 0$ **then**
    $s_k^L=1, k\in\mathcal{L}^k$;
  **elseif** $|p_{k,i}|+|q_{k,i}|=0$ **or** $|p_{k,j}|+|q_{k,j}|=0$ **then**
    $s_k^L=0, k\in\mathcal{L}^k$;
  **end if**
**end for**
$[A^{d1},\mathcal{P}_i^{d1},r_i^{d1}]=FloydProcess(\{s_k^L, k\in\mathcal{L}^k\},\{s_k^L=0, k\in\mathcal{L}^u\})$;
$[A^c,\mathcal{P}_i^c,r_i^c]=FloydProcess(\{s_k^L, k\in\mathcal{L}^k\},\{s_k^L=1, k\in\mathcal{L}^u\})$;
**for** $i\in\mathcal{N}^k$ **do**
  **if** $r_i^{d1}=0, r_i^c=1$ **then**
    $\forall k\in\mathcal{P}_i^{d1}, s_k^L=1$;
  **end if**
**end for**
**for** $i\in\mathcal{N}^k$ **do**
  **if** $r_i^c\ge 2, r_i^{d1}=0$, bus $i$ is allowed for disturbance operations, **then**
    carry out the disturbance operations;
    **for** $x\in\mathcal{G}$ **do**
      **if** $[p_{b,x}^G - p_{a,x}^G - p_i^O]=0$ **then**
        $\forall k\in\mathcal{P}_{i,x}^d, s_k^L=1$;
      **end if**
    **end for**
  **end if**
**end for**

$[A^{d2}, \mathcal{P}_i^{d2}, r_i^{d2}] = FloydProcess(\{s_k^L, k \in \mathcal{L}^k\}, \{s_k^L = 0, k \in \mathcal{L}^u\});$
**for** $i \in \mathcal{N}^k$ **do**
  **if** $r_i^{d2} = 0, s_i^N = 1$, **then**
    $\forall k \in \mathcal{F}_i, b_k^L = 2;$
  **end if**
**end for**
**output** equivalent processing results.

## IV. Case Studies

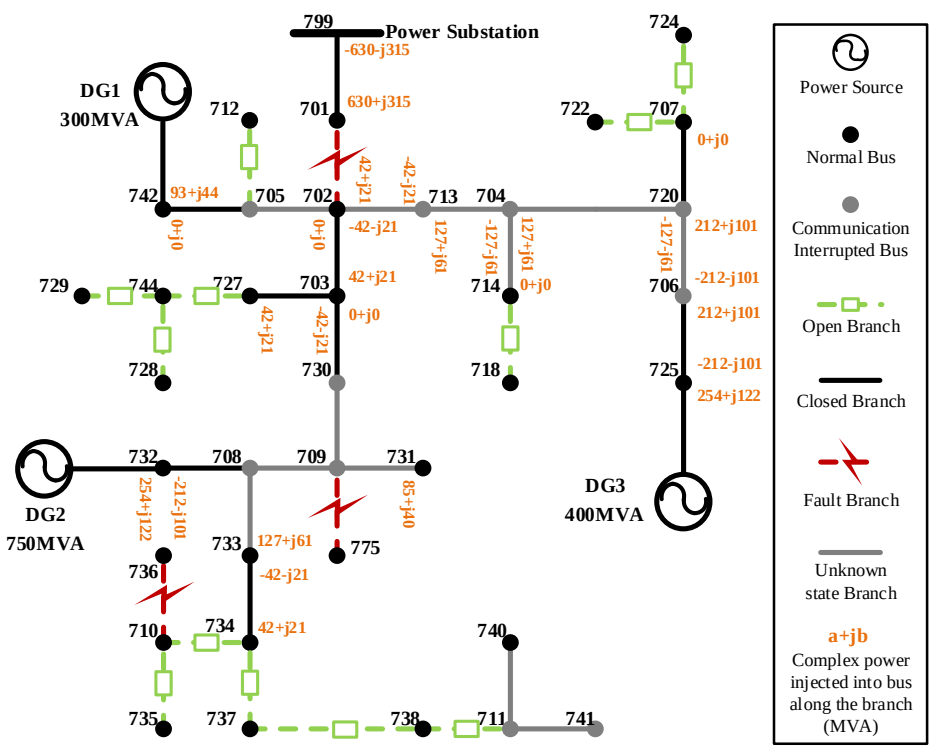

**Fig. 2.** IEEE 37-node test system with three DGs installed.

The proposed methods are validated on the IEEE 37-node test feeder system [6] with three DGs. There are 4 branches that are in failure states due to the disaster. In addition, the disaster has destroyed some communication facilities and FTUs, causing communication interruptions. The operating states of some buses and branches then become unobservable. The power flow data can be obtained from FTUs as shown in Fig. 2.

Three different processing cases are considered in verifying the effect of the proposed methods: 1) loads are not picked up during communication interruptions; 2) loads are picked up by the SCF method after Algorithm 1 is conducted; and 3) loads are picked up via the spanning tree search (STS) method.

Fig. 3(a) shows the first processing case in the test system. Its total load does not increase because there are no load recovery operations. According to the procedures in Algorithm 1, the second processing case implies that the states of branches 702-713, 713-704, 704-720, 720-706, 709-731, 708-709, and 708-733 are closed, and the states of branches 705-702, 704-714, and 730-709 are tripped, as shown in Fig. 3(c). Then, the DS restoration is carried out by the SCF method, as shown in Fig. 3(d). In the third processing case, no topology speculation is made, and loads are picked up directly via the STS method as shown in Fig. 3(b). Although the recovered load has apparently increased, $DG_2$ and $DG_3$ are overloaded and exit their operations, which further leads to a larger power outage.

By comparing the total pick-up loads of the three methods as shown in Table I and Fig. 3, the following can be concluded that: if there are no load recovery operations before communication recovery, there is no recoverable load; DGs may be overloaded if loads are picked up directly without topology inference, resulting in more load loss; and the methods proposed in this letter can effectively exploit the available information to infer the topology of DSs and help construct the restoration scheme quickly before communication recovery.

Moreover, the above methods can help construct optimized restoration schemes continuously according to the known DS component state information in the subsequent maintenance process to improve the effect of load recovery.

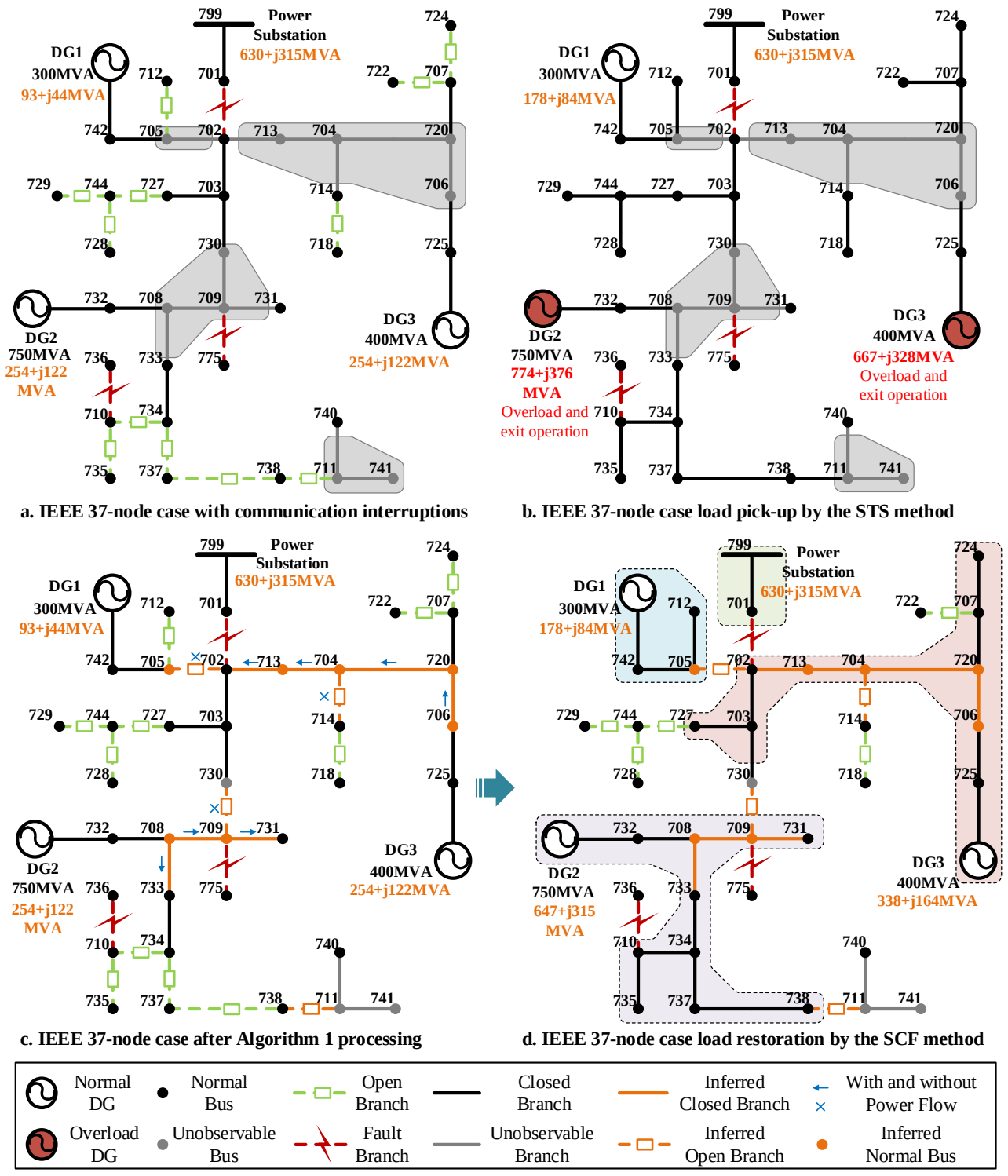

**Fig. 3.** Processing the cases in IEEE 37-node test system.

TABLE I
Picked-up Loads in Three Processing Cases

| Processing case | Total picked-up loads |
|---|---|
| 1. Do not pick up loads. | 1231+j603 MVA |
| 2. Pick up loads by the SCF method. | 1793+j878 MVA |
| 3. Pick up loads directly via the STS method. | 808+j399 MVA |

## V. Conclusion

This letter presents the three methods and corresponding algorithm for DS restoration with partial communication interruptions. The algorithm can use the available DS information to infer the states of unobservable DS components so that DS restoration can be appropriately made before communication recovery, thus improving DS resilience.